# Quaternions and Geometric Algebra
# Quaternionen und Geometrische Algebra[1]

Martin Erik Horn[2]

Email: mail@grassmann-algebra.de
mail@martin-erik-horn.de

**Abstract**

In the last one and a half centuries, the analysis of quaternions has not only led to further developments in mathematics but has also been and remains an important catalyst for the further development of theories in physics.

At the same time, Hestenes' geometric algebra provides a didactically promising instrument to model phenomena in physics mathematically and in a tangible manner. Quaternions particularly have a catchy interpretation in the context of geometric algebra which can be used didactically. The relation between quaternions and geometric algebra is presented with a view to analysing its didactical possibilities.

**Kurzfassung**

Die Untersuchung der Quaternionen führte in den vergangenen eineinhalb Jahrhunderten nicht nur zu einer Weiterentwicklung der Mathematik, sondern war und ist immer noch auch ein bedeutender Katalysator für die Weiterentwicklung physikalischer Theorien.

Gleichzeitig liefert die Geometrische Algebra von Hestenes ein didaktisch erfolgversprechendes Werkzeug, um physikalische Aussagen mathematisch anschaulich zu modellieren. Insbesondere erfahren die Quaternionen im Kontext der Geometrischen Algebra eine eingängliche Interpretation, die didaktisch genutzt werden kann. Die Beziehung zwischen dem Zahlenkörper der Quaternionen und der Geometrischen Algebra werden vorgestellt und hinsichtlich ihrer didaktischen Möglichkeiten analysiert.

**Contents**

**Inhaltsübersicht**

[1] German version published in: Volkhard Nordmeier, Arne Oberländer (Eds.): Didaktik der Physik, Beiträge zur Frühjahrstagung in Kassel, Tagungs-CD des Fachverbandes Didaktik der Physik in der Deutschen Physikalischen Gesellschaft, Beitrag 28.2, ISBN 978-3-86541-190-7 , LOB – Lehmanns Media, Berlin 2006.

[2] Address: FHW Berlin, Badensche Str. 50/51, Fach # 112, D – 10825 Berlin, Email: horn@fhw-berlin.de, University of Potsdam, Am Neuen Palais 10, D – 14469 Potsdam, Germany, Email: marhorn@uni-potsdam.de.

## 1. Introductory remarks

The theoretical description and consequent mathematisation of physical phenomenon plays a fundamental role within every didactic framework of physics (higher) education. This description however and mathematisation is not always definitive. „The world of the physical sciences is," according to Planck „a creation of the human mind that serves a certain purpose" [Planck 1952]. This means that an increasing number of competing mathematical constructs for describing physical phenomenon are possible which have been created freely by human beings.

Three of these constructs will be taken into consideration in the first part of this article: first, the quaternions discovered by Hamilton; second, the spin-matrices associated with Pauli; and third, the geometric algebra that was developed by the American physics education researcher David Hestenes.

All three mathematical approaches intersect and overlap with each other. In the second part of the article it will be attempted to reach a synthesis of these approaches using the concept of duality which will then be scrutinized within a pedagogical framework.

Physical and mathematical modelling is embedded in a temporal context just as much as cultural activities are, which can change over time. For this reason, Planck's remark [Planck 1952, p. 11] is especially relevant when he maintains that it is not uncommon that in the development of physics over time a description which had for some time been considered to be the more complicated one of two descriptions, ended up to be the more easy one as time elapsed. Even a double inversion seems to have occurred in the past: a description initially considered as easy was then discarded as too complicated only to again be regarded as easy at a later point.

While the categories "easy" and "difficult" are crucial for processes of learning, the contextual framework within which learning takes place has a much more far-reaching relevance. To use a formulation by Werner Heisenberg and Carl Friedrich von Weizsäcker: „Only theory decides what can be observed" and deducing from this: „In other words: one can only see what one (already) knows" [Weizsäcker 1994, p. 501]. Therefore, the change over from one mathematical description to another leads to the opportunity to perceive and discern something in another way. Something previously invisible can thus possibly become visible. And on the contrary, something that had previously been visible can fall into oblivion with something that is not accessible using a newly selected framework of reference.

## 1. Vorbemerkungen

Im Rahmen jeder Didaktik der (Hochschul-)Physik spielt die theoretische Beschreibung und infolgedessen die Mathematisierung physikalischer Phänomene eine wesentliche Rolle. Doch diese Beschreibung und Mathematisierung ist niemals eindeutig. „Die Welt der physikalischen Wissenschaft ist", so Planck, „eine einem bestimmten Zweck dienende Schöpfung des menschlichen Geistes" [Planck 1952]. Es werden also immer mehrere konkurrierende, frei von Menschen erschaffene mathematische Konstrukte zur Beschreibung physikalischer Phänomene möglich sein.

Drei dieser Konstrukte stehen im ersten Teil dieses Beitrags zur Diskussion: Zum ersten die von Hamilton entdeckten Quaternionen, dann die auf Pauli zurückgehenden Spin-Matrizen und drittens die vom amerikanischen Physikdidaktiker David Hestenes entwickelte Geometrische Algebra.

Diese drei mathematischen Ansätze weisen Schnittmengen und Überschneidungen auf. Ausgehend vom Konzept der Dualität wird im zweiten Teil dieses Beitrags versucht, zu einer Synthese dieser Ansätze zu gelangen, die dann didaktisch hinterfragt wird.

Wie jede Kulturtätigkeit ist auch das physikalische und mathematische Modellieren in einem zeitlichen Kontext eingebettet, der sich ändern kann. Deshalb gilt hier erst recht der Hinweis Plancks [Planck 1952, S. 11], dass es im Laufe der Entwicklung der Physik mehr als einmal vorgekommen ist, dass von zwei Beschreibungen diejenige, die eine Zeitlang als die kompliziertere galt, später als die einfachere befunden wurde. Und auch der Prozess einer doppelten Umkehrung soll schon vorgekommen sein: Die zuerst einfache Beschreibung wurde später als zu kompliziert verworfen, um dann noch später wieder als einfach zu gelten.

Doch während die Kategorien „einfach" und „kompliziert" für das Lernen entscheidend sind, kommt dem kontextuellen Rahmen, in dem diese Lernprozesse stattfinden, eine viel weitreichendere Bedeutung zu. Denn wie Werner Heisenberg und Carl Friedrich von Weizsäcker es formulieren: „Erst die Theorie entscheidet, was beobachtet werden kann" und daraus ableitend: „Anders ausgedrückt: Man sieht nur, was man weiß" [Weizsäcker 1994, S. 501]. So führt ein Wechsel eines mathematischen Beschreibungsrahmens zu einem anderen auch dazu, dass man etwas anderes sehen und erkennen kann. Etwas vorher Unsichtbares wird dann unter Umständen sichtbar. Und umgekehrt kann etwas vorher Sichtbares ins Dämmerlicht des mit dem neuen gewählten mathematischen Bezugsrahmen Unzugänglichen geschoben werden.

## 2. Quaternions

After searching for more than fifteen years with the goal in mind to theoretically adequately describe rotations in a three-dimensional space, on October 16th, 1843 William Rowan Hamilton discovered quaternions [Van der Waerden 1973]. According to legend, he received the inspiration for this three-dimensional generalisation of complex numbers during a walk with his wife along the Royal Canal in Dublin, whereupon he inscribed the fundamental relations

$$\boldsymbol{i^2 = j^2 = k^2 = ijk = -1}$$
$$\boldsymbol{ij = -ji}$$
$$\boldsymbol{jk = -kj}$$
$$\boldsymbol{ki = -ik}$$

into the Brougham bridge right then and there (Baez 2002, Conway 2003].

With this discovery Hamilton laid the foundation for a radical development not only in mathematics, but also in physics. He himself attributed great significance to his discovery and spent the rest of his life „obsessed with the quaternions” [Baez 2002].

### *Quaternions as a Calalyst of Scientific Development*

Relations that we in the present-day generally express by way of scalars and vectors were at that time formulated with the help of quaternions [Baez 2002]. Analysing the heuristic role of quaternions in physics Anderson and Joshi conclude that „the importance of the notational role of quaternions should not be underestimated“ [Anderson et al. 1992]. Thus, Maxwell initially wrote his basic equations of electrodynamics by way of components represented in the form of quaternions. Only later did Oliver Heaviside and Heinrich Hertz formulate these in the notation of vectors common today [Baylis 2002].

According to Howard Eves the discovery of Hamilton „opened the floodgates of modern abstract algebra” [Kuipers 1999, p. 4]. Modern abstract algebra continued to develop rapidly in the 20th century, and „it has been in this century that unique features of quaternionic structures have been woven closely into the development of new physical theories“ [Anderson et al. 1992].

Moreover, quaternions play a significant role in the theoretical description of real or virtual processes of all kinds of rotation. They are implemented as an important element for reducing the processing time in the field of computer graphics and are applied by default for the computation of flight navigation or in following the orbits of satellites.

## 2. Quaternionen

Nach mehr als fünfzehnjähriger Suche mit dem Ziel, Drehungen im dreidimensionalen Raum theoretisch angemessen zu beschreiben, entdeckte William Rowan Hamilton am 16. Oktober 1843 die Quaternionen [Van der Waerden 1973]. Der Legende zufolge kam ihm die Eingebung für diese dreidimensionale Verallgemeinerung der komplexen Zahlen auf einem Spaziergang mit seiner Frau entlang des Royal Canals in Dublin, worauf er die grundlegenden Beziehungen

$$\boldsymbol{i^2 = j^2 = k^2 = ijk = -1}$$
$$\boldsymbol{ij = -ji}$$
$$\boldsymbol{jk = -kj}$$
$$\boldsymbol{ki = -ik}$$

sogleich in die Brougham-Brücke ritzte [Baez 2002, Conway 2003].

Hamilton legte mit dieser Entdeckung den Grundstein für eine tiefgreifende Entwicklung nicht nur in der Mathematik, sondern auch der Physik. Er selbst maß seiner Entdeckung eine erhebliche Bedeutung zu und war den Rest seines Lebens „von Quaternionen wie besessen” [Baez 2002].

### *Quaternionen als Katalysator wissenschaftlicher Entwicklungen*

Beziehungen, die wir heute üblicherweise durch Skalare und Vektoren ausdrücken, wurden seinerzeit mit Hilfe von Quaternionen formuliert [Baez 2002]. In einer Analyse der heuristischen Rolle der Quaternionen in der Physik betonen Anderson und Joshi, dass „die Bedeutung der Quaternionen auch gerade hinsichtlich ihrer Darstellungsform nicht unterschätzt werden sollte“ [Anderson et al. 1992]. So gab Maxwell ursprünglich seine grundlegenden Gleichungen der Elektrodynamik komponentenweise in quaternionischer Form an. Erst später formulierten Oliver Heaviside und Heinrich Hertz diese in der heute üblichen Vektorschreibweise [Baylis 2002].

Hamilton öffnete mit seiner Entdeckung, so die Einschätzung von Howard Eves, „die Schleusentore der modernen abstrakten Algebra“ [Kuipers 1999, S. 4]. Im 20. Jahrhundert entwickelte sich diese rasant weiter, und es war in diesem Jahrhundert, „dass wesentliche und einzigartige Eigenschaften quaternionischer Strukturen eng mit der Entwicklung moderner physikalischer Theorien verwoben wurden“ [Anderson et al. 1992].

Darüber hinaus spielen Quaternionen bei der theoretischen Beschreibung realer oder virtueller Rotationsprozesse aller Art eine bedeutende Rolle. In der Computergraphik werden sie als wichtiges Element zur Reduzierung der Rechenzeit eingesetzt und standardmäßig bei Berechnungen zur Flugnavigation oder der Bahnverfolgung von Satelliten verwendet.

The mathematics involved in quaternions is apparently extremely application-oriented so that the US Air Force has financed the publication of textbooks on this subject [Kuipers 1999]. And the mathematics of the quaternions is apparently so easy to understand, as is mentioned in the preface of the mentioned textbook, „that much of the subject matter would be accessible to those with a modest background in mathematics“ [Kuipers 1999, p. XIX]. This accessibility militates in favour of the didactic strength of theories formulated with quaternions.

***„Quaternion Ambiguity“***

Hamilton's approach was however not received without causing controversy. Following a partially rancorous debate amongst adherents of quaternions and adherents of vector algebra, the mathematical system which is more common today, the – according to Doran and Lasenby – „hybrid system of vector algebra“ [Doran et al. 2003, p. 10] finally succeeded.

The reverberations of this inner-physical and inner-mathematical trench warfare are still felt today. A conflicting and ambivalent treatment of quaternions can be observed in the numerous representations of basic mathematical principles in physics. For instance, the illustration of the basic mathematical principles of the Dirac equation is a typical example of this "quaternion ambiguity".

In this way Penrose explicitly characterises the formulation of the relativistic equation of the description of electrons by Dirac as a "rediscovery"[3]. He emphasises the brilliance of the Dirac formulation while however simultaneously noting that already William Rowan Hamilton had specified the "square root" of the three-dimensional Laplacian by means of quaternions as

$$\left(i\frac{\partial}{\partial x}+j\frac{\partial}{\partial y}+k\frac{\partial}{\partial z}\right)^2=-\left(\frac{\partial}{\partial x}\right)^2-\left(\frac{\partial}{\partial y}\right)^2-\left(\frac{\partial}{\partial z}\right)^2=-\nabla^2.$$

This quaternionic relationship constitutes the foundation for the Dirac equation [Penrose 2005, p. 619]. A more emphatic substantiation for the relevance and significance of quaternions can hardly be given.

Even historically the significance of quaternions within the context of the Dirac equation was analysed shortly after its release. Already Arnold Sommerfeld submitted a scientific contribution of Walter Franz [Franz 1935] about the methodology of the quaternionic Dirac equations to the Bavarian Aca-

[3] As Penrose wrote, „Dirac rediscovered (an instance of) the C*lifford algebras*“ [Penrose 2005, p. 619].

Die Mathematik der Quaternionen ist offensichtlich so anwendungsorientiert, dass die US Air Force entsprechende Lehrbücher [Kuipers 1999] finanziert. Und die Mathematik der Quaternionen ist offensichtlich so leicht zu verstehen, dass diese auch dem Personal der US Air Force mit einem, wie es im Vorwort des besagten Werkes höflich heißt, „moderaten Hintergrund“ an mathematischem Wissen zugänglich ist [Kuipers 1999, S. XIX]. Diese leichte Zugänglichkeit spricht für die didaktische Stärke quaternionisch formulierter Theorien.

***„Quaternionen-Ambiguität“***

Hamiltons Ansatz war jedoch nicht unumstritten. In einer teilweise erbittert geführten Auseinandersetzung zwischen Anhängern der Quaternionen und Anhängern der heute üblichen Vektoralgebra setzte sich das nach Einschätzung von Doran und Lasenby „hybride System der Vektoralgebra“ („...by the hybrid system of vector algebra.“) [Doran et al. 2003, S. 10] schließlich durch.

Der Nachhall dieser inner-physikalischen und inner-mathematischen Grabenkämpfe ist auch heute noch zu spüren. In zahlreichen Darstellungen der mathematischen Grundlagen der Physik zeigt sich ein überaus zwiespältiger und ambivalenter Umgang mit den Quaternionen. Ein typisches Beispiel für diese „Quaternionen-Ambiguität“ findet sich beispielsweise bei der Darstellung der mathematischen Grundlagen der Dirac-Gleichung.

So bezeichnet Penrose die Formulierung der relativistischen Gleichung zur Beschreibung des Elektrons durch Dirac ausdrücklich als „Wiederentdeckung“[3]. Er hebt die Brillanz der Dirac’schen Formulierung hervor, weist jedoch gleichzeitig darauf hin, dass schon William Rowan Hamilton die „Quadratwurzel“ des dreidimensionalen Laplace-Operators mit Hilfe der Quaternionen als

$$\left(i\frac{\partial}{\partial x}+j\frac{\partial}{\partial y}+k\frac{\partial}{\partial z}\right)^2=-\left(\frac{\partial}{\partial x}\right)^2-\left(\frac{\partial}{\partial y}\right)^2-\left(\frac{\partial}{\partial z}\right)^2=-\nabla^2$$

angegeben habe. Diese quaternionische Beziehung bildet die Grundlage der Dirac-Gleichung [Penrose 2005, S. 619]. Eine nachdrücklichere Begründung für die Bedeutung der Quaternionen kann kaum gegeben werden.

Auch historisch wurde die Bedeutung der Quaternionen im Kontext der Dirac-Gleichungen schon kurz nach deren Veröffentlichung untersucht. So ist in einer durch Arnold Sommerfeld bei der Bayerischen Akademie der Wissenschaften eingereichten Abhandlung von Walter Franz zur Methodik der Dirac-demy of Sciences. There Franz analysed a formulation of the Dirac equation with the help of complex quaternions. „The most obvious advantage", explains Franz this quaternionic approach, „lies in the conceptual simplicity and clarity of the method that for instance allows an especially simple proof of the Lorentz invariance of equations" [Franz 1935, p. 381].

[3] Penrose schreibt wörtlich: „Dirac entdeckte (ein Beispiel der) C*lifford-Algebras* wieder“ [Penrose 2005, S. 619].

Despite numerous advantages and a diversity of possibilities for application [Gsponer et al. 2005], quaternions are for the large part ignored in current physics textbooks at the university level today. These discrepancies between the significance and representation or illustration of quaternions is evidence of an ambivalent and highly conflictual treatment that might unconsciously reflect previous debates and disputes.

And also Penrose, who emphasises the significance of quaternions for understanding the Dirac equation, smiles at Hamilton in another part of his book: „However, from our present perspective, as we look back over the $19^{th}$ and $20^{th}$ centuries, we must still regard these heroic efforts as having resulted in relative failure" [Penrose 2005, p. 200].

***The Significance of Quaternions in Education***

In addition to the clarification of conceptual foundations of physics and the already mentioned ease in comprehension, quaternions exhibit further dimensions of significance that should not be underestimated in their relevance for education.

One of the most important ones lies in the capability of quaternions to provide new metaconceptual views that present themselves with the generalisation of quaternionic approaches. Anderson and Joshi describe, that „the extra structure of quaternions over complex and real numbers has enabled new perspectives through permitting different formalisms."

And what is more important: These new perspectives „provided structures within which new physical theories can be considered" [Anderson et al. 1992, p. 20].

The development of new perspectives can thus support processes of conceptual change in two respects. On the one hand, historic changes in concepts can be traced and analysed. On the other, changes in concepts can be attained in the learning process which enable students to gain a superior perspective. An understanding of physical phenomenon requires that the learners not only take on different perspectives and standpoints, but that they are also able to question them in a metaconceptual manner [Horn 2003].

Gleichung [Franz 1935] diese mit Hilfe von komplexen Quaternionen formuliert. „Der offensichtlichste Vorteil", erläutert Franz diesen quaternionischen Ansatz, „liegt in der begrifflichen Einfachheit und Übersichtlichkeit der Methode, die z.B. einen besonders einfachen Beweis für die Lorentz-Invarianz der Gleichungen gestattet." [Franz 1935, S. 381].

Trotz zahlreicher weiterer Vorzüge und einer Vielzahl von Anwendungsmöglichkeiten [Gsponer et al. 2005] werden Quaternionen in den heute verbreiteten universitären Physik-Lehrbüchern zumeist übergangen. Diese Diskrepanz zwischen Bedeutung und Darstellung der Quaternionen zeugt von einem ambivalenten und höchst zwiespältigen Umgang, der unbewusst die vergangenen Auseinandersetzungen widerspiegeln mag.

Und auch Penrose, der die Bedeutung der Quaternionen für ein Verständnis der Dirac-Gleichung hervorhebt, mokiert sich an anderer Stelle seines Buches [Penrose 2005, S. 200] über „diese heldenhaften Versuche" Hamiltons, reine Quaternionen zur Modellierung physikalischer Theorien heranzuziehen, die „alle relativ gescheitert seien."

***Didaktische Bedeutung der Quaternionen***

Neben der Klärung konzeptioneller Grundlagen der Physik und der schon erwähnten leichten Zugänglichkeit kommen den Quaternionen noch weitere Bedeutungsdimensionen zu, die ebenfalls eine nicht zu unterschätzende didaktische Relevanz aufweisen.

Eine der wichtigsten ist dabei die Ermöglichung metakonzeptueller Sichtweisen, die sich bei Verallgemeinerung quaternionischer Ansätze bieten. Anderson und Joshi beschreiben, dass „die zusätzlichen Strukturen reeller und komplexer Quaternionen neue Perspektiven ermöglicht haben, da verschiedenartige Formalismen zugelassen werden."

Und weit wichtiger noch: „Sie lieferten darüber hinaus Strukturen, mit deren Hilfe neue physikalische Theorien betrachtet werden können" [Anderson et al. 1992, S. 20].

Die Herausbildung neuer Perspektiven kann somit Conceptual-Change-Prozesse in zweierlei Hinsicht unterstützen. Einerseits lassen sich historische Konzeptwechsel nachverfolgen und analysieren. Andererseits werden im Lernprozess Konzeptwechsel ermöglicht, die bei den Studenten zu einer übergeordneten Sichtweise führen. Denn ein Verständnis physikalischer Phänomene setzt voraus, dass die Lernenden nicht nur unterschiedliche Blickrichtungen und Standpunkte einnehmen, sondern diese auch metakonzeptuell hinterfragen können [Horn 2003].

## 3. Pauli matrices

Wolfgang Pauli during his lifetime, according to the Constance physics historian Ernst-Peter Fischer, was „at the border of thought“. In his biography of Pauli [Fischer 2000], Fischer especially illustrates the psychological background of Pauli's struggle with an integrated and holistic description of the world.

One central aspect according to Fischer is the complex psychological transition from the medieval trinity (back) to the Pythagorean quantum mechanical quaternity, which according to Pauli enables „a greater completeness of experience” [Pauli 1994, p. 258].

All the more surprising is thus that Pauli draws upon the trivalent matrices

$$\sigma_x^2 = \sigma_y^2 = \sigma_z^2 = \mathbf{1}$$
$$\sigma_x\sigma_y = -\sigma_y\sigma_x$$
$$\sigma_y\sigma_z = -\sigma_z\sigma_y$$
$$\sigma_z\sigma_x = -\sigma_x\sigma_z$$

for describing the nonrelativistic electron. The derivation of these matrices, a similar version of which can be found in all newer physics textbooks, is represented in a very appealing manner in the quantum mechanics lecture given by Enrico Fermi [Fermi 1995] and in the concise depiction by Sin-itiro Tomonaga [Tomonaga 1997].

### *The Algebraic Relationship between Quaternions and Pauli Matrices*

In accordance with a portrayal by Abraham Pais [Pais 2002, p.209], already shortly after the formulation of the spin matrices by Pauli, Ernst Pascual Jordan calls Pauli's attention to the close relationship between quaternions and the newly discovered spin matrices. This relationship is written by Élie Cartan[4] [Cartan 1981, p. 45] as:

$$\boldsymbol{i} = -\,\mathrm{i}\,\sigma_x = -\,\mathrm{i}\begin{pmatrix} 0 & 1 \\ 1 & 0 \end{pmatrix}$$

$$\boldsymbol{j} = -\,\mathrm{i}\,\sigma_y = -\,\mathrm{i}\begin{pmatrix} 0 & -\mathrm{i} \\ \mathrm{i} & 0 \end{pmatrix}$$

$$\boldsymbol{k} = -\,\mathrm{i}\,\sigma_z = -\,\mathrm{i}\begin{pmatrix} 1 & 0 \\ 0 & -1 \end{pmatrix}$$

## 3. Pauli-Matrizen

Wolfgang Pauli bewegte sich, so der Konstanzer Physikhistoriker Ernst-Peter Fischer, Zeit seines Lebens „an den Grenzen des Denkens“. Fischer stellt in seiner Biographie Paulis [Fischer 2000] gerade auch die psychologischen Hintergründe von Paulis Ringen mit einer ganzheitlichen Weltbeschreibung dar.

Ein zentraler Punkt ist dabei nach Fischer der psychologisch nicht einfache Übergang von der mittelalterlichen Trinität (zurück) zur pythagoräisch-quantenmechanischen Quaternität, die nach Pauli eine „größere Vollständigkeit des Erlebens“ [Pauli 1994, S. 258] verspricht.

Umso erstaunlicher ist deshalb, dass Pauli zur Beschreibung des nichtrelativistischen Elektrons die dreiwertigen Matrizen

$$\sigma_x^2 = \sigma_y^2 = \sigma_z^2 = \mathbf{1}$$
$$\sigma_x\sigma_y = -\sigma_y\sigma_x$$
$$\sigma_y\sigma_z = -\sigma_z\sigma_y$$
$$\sigma_z\sigma_x = -\sigma_x\sigma_z$$

heranzieht. Die Herleitung dieser Matrizen, die sich in ähnlicher Form auch in allen neueren Physik-Lehrbüchern findet, ist sehr ansprechend in der Quantenmechanik-Vorlesung von Enrico Fermi [Fermi 1995] und der übersichtlichen Darstellung von Sin-itiro Tomonaga [Tomonaga 1997] dargelegt.

### *Die algebraische Beziehung zwischen Quaternionen und Pauli-Matrizen*

Schon kurz nach Formulierung der Spin-Matrizen durch Pauli, wird er – einer Schilderung von Abraham Pais zufolge [Pais 2002, S. 290] – von Ernst Pascual Jordan auf die enge Beziehung zwischen den Quaternionen und den neu entdeckten Spin-Matrizen aufmerksam gemacht. Diese Beziehung lautet in der Schreibweise[4] von Élie Cartan [Cartan 1981, S. 45]:

$$\boldsymbol{i} = -\,\mathrm{i}\,\sigma_x = -\,\mathrm{i}\begin{pmatrix} 0 & 1 \\ 1 & 0 \end{pmatrix}$$

$$\boldsymbol{j} = -\,\mathrm{i}\,\sigma_y = -\,\mathrm{i}\begin{pmatrix} 0 & -\mathrm{i} \\ \mathrm{i} & 0 \end{pmatrix}$$

$$\boldsymbol{k} = -\,\mathrm{i}\,\sigma_z = -\,\mathrm{i}\begin{pmatrix} 1 & 0 \\ 0 & -1 \end{pmatrix}$$

---

[4] In some books (see for instance [Doran et al. 2003, p. 34] ) one can find another use of signs, so that one has to differentiate between an illustration of a right-handed system of coordinates on a left-handed one. By means of one or three changes of signs these different illustrations can be transferred into each other.

[4] In manchen Büchern [z.B. Doran et al. 2003, S. 34] findet sich eine andere Vorzeichenwahl ($\boldsymbol{i} = \mathrm{i}\,\sigma_x$, etc...), so dass zwischen der Darstellung in einem rechts- und einem linkshändigen Koordinatensystem unterschieden werden muss. Durch ein oder drei Vorzeichenwechsel können diese unterschiedlichen Darstellungen ineinander überführt werden.

Special attention should be paid to the difference between the complex base unit **i** and the quaternion base unit ***i*** which both have different meanings and different functions.

How these different meanings and functions should be understood depends on the interpretation. As long as no interpretation is undertaken, the relationship is of a purely algebraic nature. In the manner of speaking common to Weizsäcker, one would say that „physical semantics are missing" [Weizsäcker 1994, p. 501].

### *The Meaning of the Pauli Matrices*

The mathematical takeover or „assumption of power of mathematics in quantum theory", which according to Carl Friedrich von Weizsäcker dates back to the year 1932 [Weizsäcker 1994, p. 511], has long-term consequences for the interpretation of all mathematical structures connected with quantum mechanics. Since this time there exists a strictly axiomatic structure of quantum mechanics. And therefore it appears at times as if the meaning of fundamental parts of this theory has been like frozen into the ice of mathematics since this unfriendly takeover.

This in effect is true for the interpretation of the Pauli matrices. Since their introduction by Pauli, they have been derived from, been described and interpreted exclusively as quantum mechanic operators in nearly all textbooks. No alternative approaches of explanation are neither attempted nor given. Thus, the impression that Pauli matrices are something absolutely quantum mechanically is automatically generated in the teaching and learning process of students.

Thereby, the description of spin matrices as operators is very similar to the description of complex numbers – and these can be interpreted in two completely different contexts.

### *The Complementary Interpretation of Complex Numbers*

The complementary interpretation of complex numbers is clearly evident in the Gaussian plane. On the one hand, complex numbers are interpreted as coordinates. Thus, the complex number **z = 5 + 3 i** is represented by a vector with five basic units in real direction and three basic units in imaginary direction (see figure 1).

The vectors represented in the Gaussian plane can be manipulated with simple multiplications: a multiplication with a real number causes a compression or a dilation. A multiplication with a complex number of norm 1 causes a rotation, and a multiplication with an arbitrary complex number causes a rotation that is connected with a compression or a dilation.

Zu achten ist dabei insbesondere auf den Unterschied zwischen der komplexen Basiseinheit **i** und der quaternionischen Basiseinheit ***i***, die eine unterschiedliche Bedeutung und eine unterschiedliche Funktion haben.

Wie diese unterschiedliche Bedeutung und Funktion zu verstehen ist, ist jedoch eine Frage der Interpretation. So lange, wie diese Interpretation nicht vorgenommen wird, handelt es sich um eine Beziehung rein algebraischer Natur. In der Sprechweise von Weizsäckers müsste man sagen: „Es fehlt die physikalische Semantik." [Weizsäcker 1994, S. 501].

### *Die Bedeutung der Pauli-Matrizen*

Die „Machtübernahme der Mathematik in der Quantentheorie", die nach Carl Friedrich von Weizsäcker in das Jahr 1932 fällt [Weizsäcker 1994, S. 511], hat für die Deutung der mit ihr verbundenen mathematischen Strukturen langfristige Konsequenzen. Es ist nun ein streng axiomatischer Aufbau vorhanden, und so hat es teilweise den Anschein, dass die Bedeutung grundlegender Teile dieser Theorie seit dieser „Machtübernahme" wie eingefroren festliegen.

In der Tat gilt dies für die Interpretation der Pauli-Matrizen. Sie werden seit Einführung durch Pauli in nahezu allen Lehrbüchern ausschließlich als quantenmechanische Operatoren hergeleitet, beschrieben und interpretiert. Alternative Erklärungsansätze werden nicht versucht und nicht gegeben. Bei den Lernenden muss dadurch der Eindruck entstehen, Pauli-Matrizen seien etwas originär Quantenmechanisches.

Dabei ähnelt die Beschreibung der Spinmatrizen als Operatoren sehr der Beschreibung von komplexen Zahlen – und diese lassen sich in zwei gänzlich unterschiedlichen Bedeutungszusammenhängen interpretieren.

### *Die komplementäre Deutung der komplexen Zahlen*

Die komplementäre Deutung der komplexen Zahlen wird in der Gauß'schen Zahlenebene anschaulich sichtbar. Einerseits werden komplexe Zahlen im Sinne von Koordinaten gedeutet. So entspricht die komplexe Zahl **z = 5 + 3 i** einem Vektor, der fünf Basiseinheiten in reeller Richtung und drei Basiseinheiten in imaginärer Richtung zeigt (siehe Abbildung 1).

Die in der Gauß'schen Zahlenebene dargestellten Vektoren lassen sich durch einfache Multiplikationen manipulieren: Eine Multiplikation mit einer reellen Zahl bewirkt eine Stauchung oder Streckung. Eine Multiplikation mit einer komplexen Zahl der Norm 1 bewirkt eine Drehung, und eine Multiplikation mit einer beliebigen komplexen Zahl bewirkt

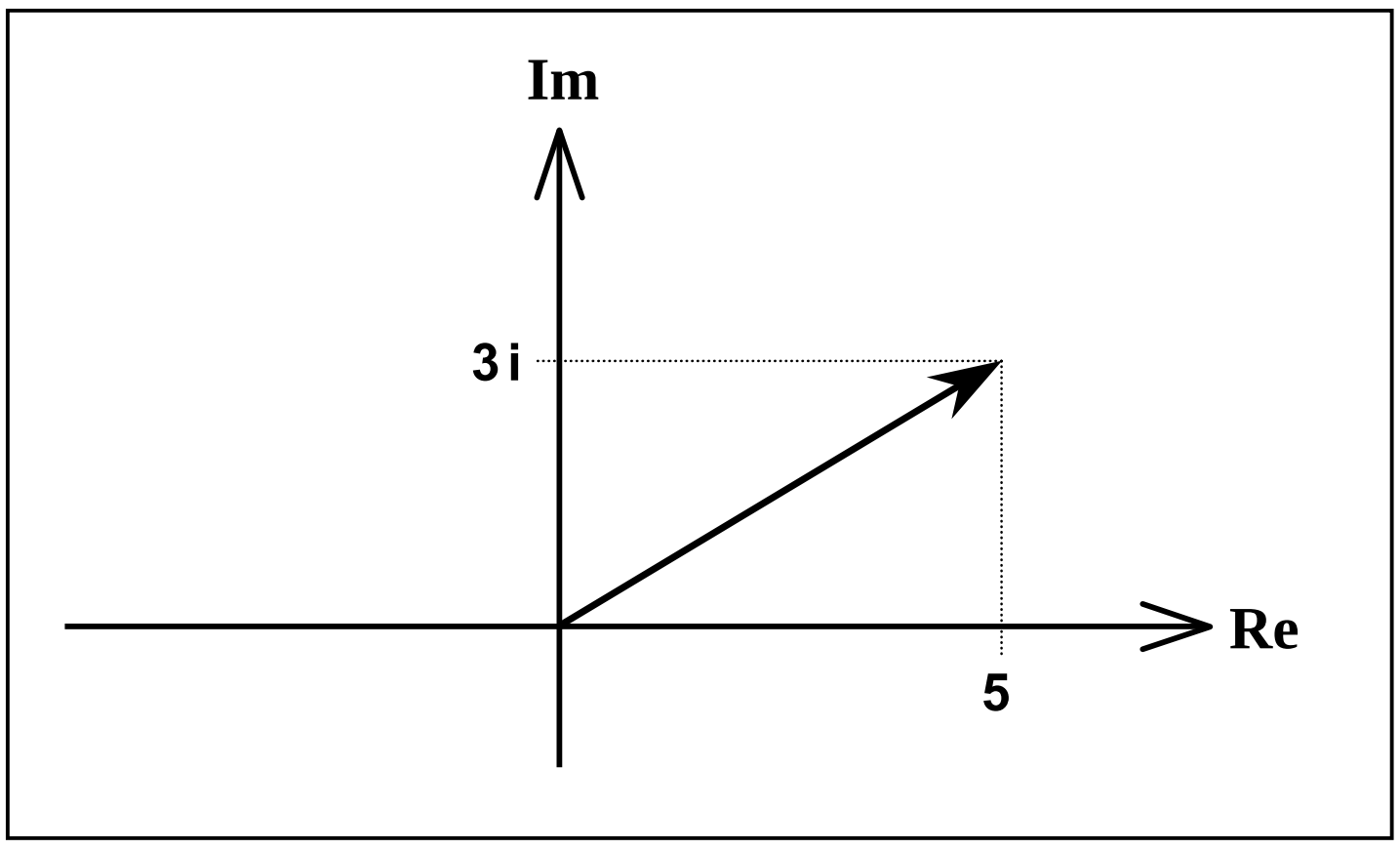

Fig. 1: Representation of the complex number **z = 5 + 3 i** in the complex plane.

Abb. 1: Darstellung der komplexen Zahl **z = 5 + 3 i** in der Gauß'schen Zahlenebene.

Complex numbers thus have an operational character. The number **i** for example imparts a rotation of 90° (see figure 2). Complex numbers can consequently be interpreted as **coordinates** as well as **operators**. This complementary perspective is likewise possible with the Pauli matrices. Pauli matrices are not mere operators, but also – coordinates!

This finding is at the centre of geometric algebra: Pauli matrices can also be interpreted as coordinates of our three-dimensional, Euclidean space (within which we lived before Einstein's discovery of the relativity theory).

eine Drehung, die mit einer Stauchung oder Streckung verbunden ist.

Komplexe Zahlen haben somit einen operationalen Charakter. Die Zahl **i** vermittelt beispielsweise eine Rotation um 90° (siehe Abbildung 2). Somit können komplexe Zahlen sowohl als **Koordinaten** wie auch als **Operatoren** gedeutet werden. Diese komplementäre Sichtweise ist auch bei den Pauli-Matrizen möglich. Pauli-Matrizen sind eben nicht nur Operatoren, sondern auch – Koordinaten!

Im Mittelpunkt der Geometrischen Algebra steht genau diese Erkenntnis: Pauli-Matrizen können **auch** als Koordinaten unseres dreidimensionalen, euklidischen Raumes (in dem wir vor Einsteins Entdeckung der Relativitätstheorie lebten) interpretiert werden.

Fig. 2: Representation of the complex number **i** as rotation operator in the complex plane.

Abb. 2: Darstellung der komplexen Zahl **i** als Drehoperator in der Gauß-schen Zahlenebene.

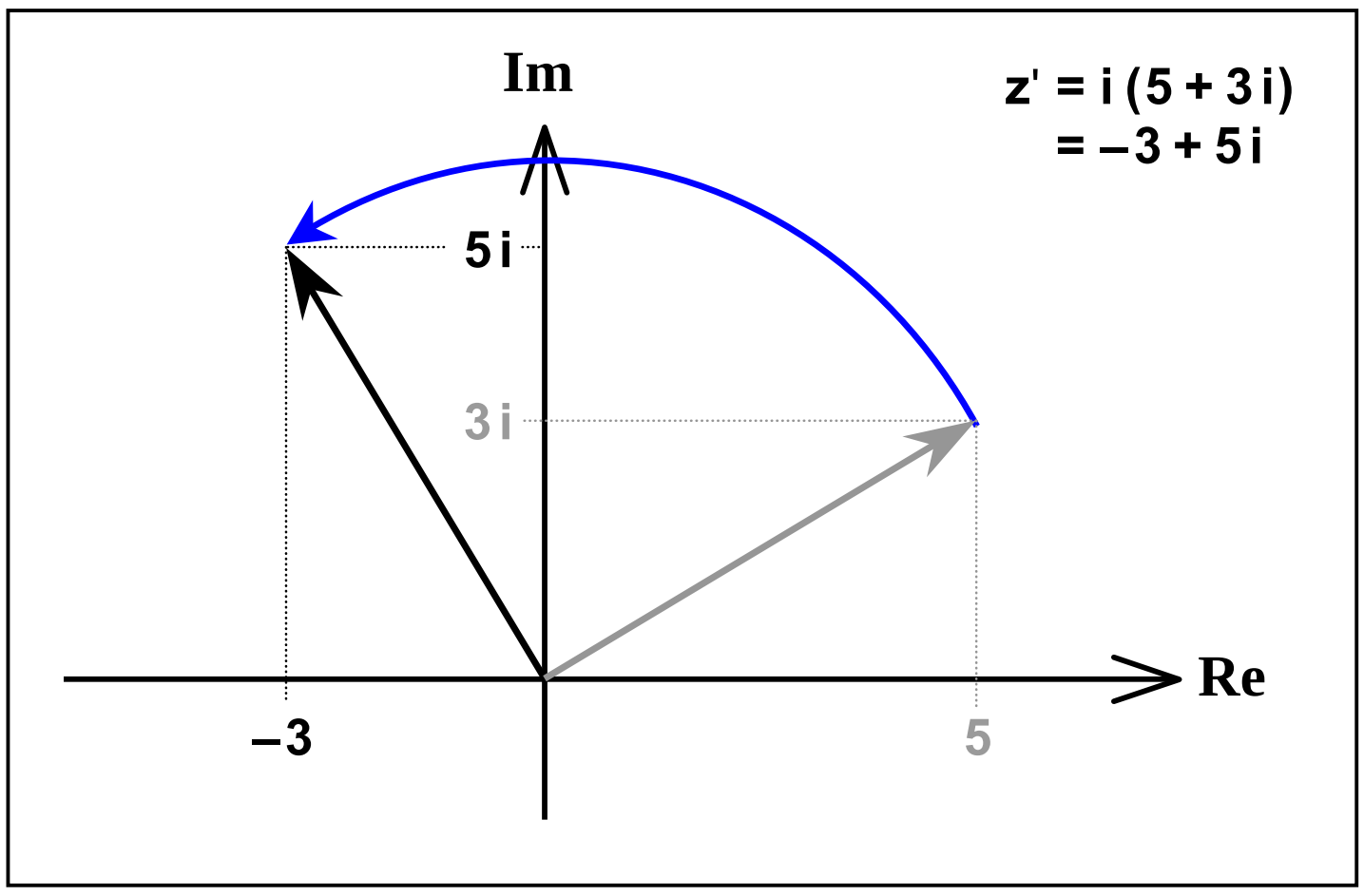

### 4. Geometric algebra

Geometric algebra which for the first time was formulated by the American physics education researcher David Hestenes ( see [Hestenes 1990], [Hestenes 2002], [Doran 2003] and the references listed therein) provides a mathematical language that is adequate in comprehensively formulating physical phenomenon in a mathematical consistent manner [Hestenes 1986].

### 4. Geometrische Algebra

Die erstmalig durch den amerikanischen Physikdidaktiker David Hestenes formulierte Geometrische Algebra ( siehe [Hestenes 1990], [Hestenes 2002], [Doran 2003] und die darin aufgeführten Literaturhinweise) liefert eine mathematische Sprache, die geeignet ist, physikalische Phänomene übergreifend mathematisch zu formulieren [Hestenes 1986].

At the university, students of physics are currently confronted with a diversity of competing mathematical approaches. These approaches however are not constructively interrelated in terms of a metaconceptual perspective, but rather make up a truly Babylonian chaos of different mathematical speech that, according to David Hestenes, leads to massive redundancies in the language or languages we are using in mathematics: „Suffering from a Babylon of tongues, mathematics has become massively redundant“ [Hestenes 1986, p. 3].

Geometric algebra is trying to recognise these redundancies by way of a metaconceptual integration of a diversity of mathematical dialects and diffuse the problem by the means of education and methods of teaching.

This can be achieved by integrating the so-called „geometric product“ already formulated by Hermann Günther Graßmann and more closely analysed by William Kingdon Clifford which demonstrates a geometrically founded unification of the inner and outer product. The approach by Hestenes distinguishes itself in that algebraic and geometric interpretations are associated with each other stringently and complementarily.

Während des Studiums werden Physikstudentinnen und -studenten derzeit mit einer Vielzahl konkurrierender mathematischer Ansätze konfrontiert. Diese Ansätze werden dabei nicht konstruktiv im Sinne einer metakonzeptuellen Sichtweise verknüpft, sondern bilden nach David Hestenes ein wahrlich „babylonisches Chaos an mathematischen Sprechweisen“, das dazu führt, dass die derzeit von uns benutze Sprache bzw. benutzten Sprachen in der Mathematik massive Redundanzen aufweisen. [Hestenes 1986, S. 3].

Die Geometrische Algebra versucht, diese Redundanzen durch die metakonzeptuelle Vernetzung der unterschiedlichen mathematischen Dialekte zu erkennen und didaktisch zu entschärfen.

Dies gelingt durch die Einbindung des bereits von Hermann Günther Graßmann formulierten und von William Kingdon Clifford näher untersuchtem sogenannten „geometrischen Produkts“, das eine geometrisch begründete Vereinheitlichung von innerem und äußerem Produkt darstellt. Dieser Hestenes’sche Ansatz zeichnet sich insbesondere dadurch aus, dass algebraische und geometrische Deutungen in einen stringenten und sich komplementär ergänzenden Zusammenhang stehen.

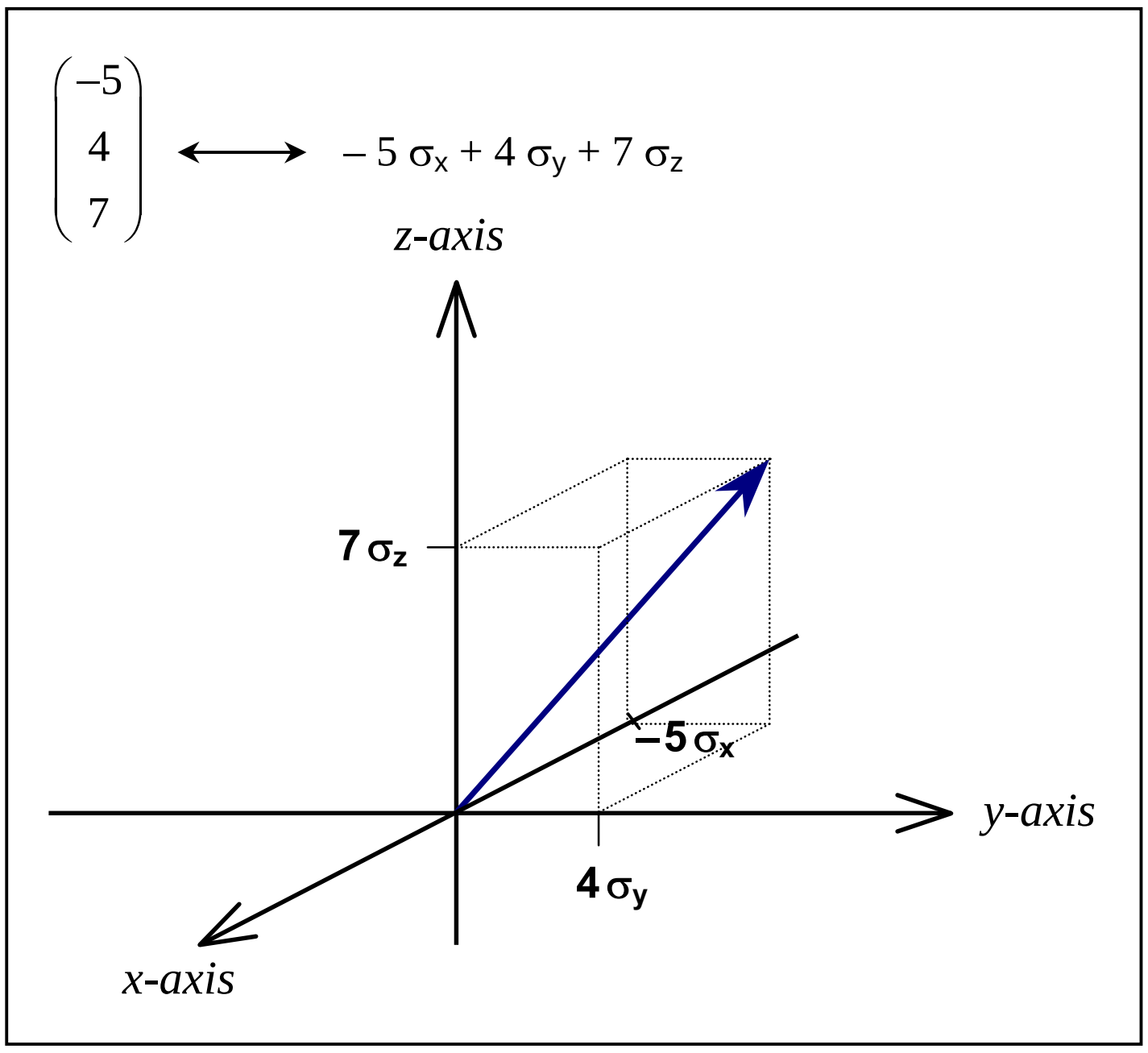

Fig. 3: Representation of a vector in geometric algebra.

Abb. 3: Darstellung eines Vektors in der Geometrischen Algebra.

At the centre of this interrelation of algebra and geometry is the interpretation of the Pauli matrices as basic vectors of the three-dimensional, Euclidean space. Vectors are consequently depicted in geometric algebra as a linear combination of the three Pauli matrices.

Im Zentrum dieser Verknüpfung von Algebra und Geometrie steht die Deutung der Pauli-Matrizen als Basisvektoren des dreidimensionalen, euklidischen Raumes (siehe Abbildung 3). Vektoren werden in der Geometrischen Algebra deshalb als Linearkombination der drei Pauli-Matrizen dargestellt.

### *The Basis of Geometric Algebra*

The geometric interpretation of Pauli matrices becomes more clear when considering the products of Pauli matrices. The product of two different Pauli matrices can be interpreted as an area element. Any plane can therefore be represented as a linear combination of the three possible directed area elements $\sigma_x\sigma_y$, $\sigma_y\sigma_z$ and $\sigma_z\sigma_x$. Within this context bivectors experience a natural interpretation as directed planes, while trivectors as a multiple of the volume element $\sigma_x\sigma_y\sigma_z$ can be interpreted as directed volumes (see figure 4).

| mathematical object | | geometric interpretation |
|---|---|---|
| scalar | **1** | dimensionless number |
| vectors | $\sigma_x$ $\sigma_z$ $\sigma_y$ | directed lines |
| bivectors *(pseudo-vectors)* | $\sigma_x\sigma_y$ $\sigma_y\sigma_z$ $\sigma_z\sigma_x$ | directed areas |
| trivector *(pseudoscalar)* | $\sigma_x\sigma_y\sigma_z$ | directed volume |

Fig. 4: Basic elements of geometric algebra.

In order to mathematically model processes in a three-dimensional Euclidean space, a total of eight basic units are required. The meaning and interpretation of these mathematical parameters are at the centre of attention here.

### *Forms of Representation of Geometric Algebra*

The meaning of mathematical constructs however is always imparted by the relationships of these constructs to other mathematical constructs and never by the representation of mathematical constructs as such (see for instance [Jeans 1981] ). Baylis therefore concludes that „we can concentrate on the algebra and forget explicit representations” when working with Pauli matrices [Baylis 2002, p. 12].

Even from a didactic point of view, the representation of matrices is problematic: „Working directly with matrices does obscure geometric meaning, and is usually best avoided“ [Doran et al. 2003, p. 38]. With the above-mentioned formulae

$$\boldsymbol{i} = -\,\mathbf{i}\,\sigma_x = -\,\mathbf{i}\begin{pmatrix}0 & 1\\ 1 & 0\end{pmatrix},\quad \text{etc...}$$

### *Die Basisgrößen der Geometrischen Algebra*

Die geometrische Interpretation der Pauli-Matrizen wird noch deutlicher, wenn die Produkte der Pauli-Matrizen betrachtet werden. Das Produkt aus zwei unterschiedlichen Pauli-Matrizen kann als flächenhafte Basiseinheit gedeutet werden. Jede beliebige Fläche lässt sich deshalb als Linearkombination der drei möglichen Basisgrößen $\sigma_x\sigma_y$, $\sigma_y\sigma_z$ und $\sigma_z\sigma_x$ darstellen. Bivektoren erfahren in diesem Kontext eine natürliche Deutung als orientierte Flächen, während Trivektoren als Vielfaches der Basiseinheit $\sigma_x\sigma_y\sigma_z$ als Volumina interpretiert werden können (siehe Abbildung 4).

| Mathematisches Objekt | | Geometrische Interpretation |
|---|---|---|
| Skalar | **1** | dimensionslose Größe |
| Vektoren | $\sigma_x$ $\sigma_z$ $\sigma_y$ | gerichtete Strecken |
| Bivektoren *(Pseudovektoren)* | $\sigma_x\sigma_y$ $\sigma_y\sigma_z$ $\sigma_z\sigma_x$ | orientierte Flächen |
| Trivektor *(Pseudoskalar)* | $\sigma_x\sigma_y\sigma_z$ | orientiertes Volumen |

Abb. 4: Basisgrößen der Geometrischen Algebra.

Um Vorgänge in einem dreidimensionalen euklidischen Raum mathematisch zu modellieren, werden insgesamt acht Basisgrößen benötigt. Dabei steht die Bedeutung und Interpretation dieser mathematischen Größen im Zentrum.

### *Repräsentationsformen der Geometrischen Algebra*

Die Bedeutung mathematischer Konstrukte wird jedoch immer durch die Beziehung dieser Konstrukte zu anderen mathematischen Konstrukten vermittelt und nie durch die Darstellung mathematischer Konstrukte an sich (siehe zum Beispiel [Jeans 1981] ). Baylis schlussfolgert deshalb, dass „wir uns auf die Algebra der Pauli-Matrizen an sich konzentrieren sollten, und explizite Repräsentationen als Matrizen vergessen können“ [Baylis 2002, S. 12].

Auch aus didaktischer Sicht ist die Matrizendarstellung problematisch: „Direkt mit 2x2-Matrizen zu arbeiten, verschleiert die geometrische Bedeutung und wird am besten gleich vermieden“ [Doran et al. 2003, S. 38]. An den oben angegebenen Formeln

$$\boldsymbol{i} = -\,\mathbf{i}\,\sigma_x = -\,\mathbf{i}\begin{pmatrix}0 & 1\\ 1 & 0\end{pmatrix},\quad \text{etc...}$$

the equalisation with a 2x2 representation of matrices is not what is most important in this regard, but rather the relationship between quaternions and Pauli matrices formulated at the left side of these equations. They thus show the relationship between two different mathematical constructs.

### *Teaching Geometric Algebra*

There are a great number of possible connecting points for developing didactical theories for teaching geometric algebra. In the usual representation of geometric algebra, the emphasis is placed on beginning with the interpretation of the geometric product $\mathbf{ab} = \mathbf{a} \cdot \mathbf{b} + \mathbf{a} \wedge \mathbf{b}$.

This approach is based on the fact that within the framework of vector algebra in use today, the product of two vectors continues to be defined in an inconsistent and unclear manner. As a matter of fact, with the scalar product and the cross product (vector product) there are two different products of vectors that exist which are presented in textbooks completely separate from each other. In addition the cross product does not allow an unambiguous and definite generalisation in higher-dimensional spaces.

Then learners are at first expected to „unlearn“ previous knowledge („a little un-learning“)[5] [Gull et al. 1993]. The hitherto existing knowledge of students is labeled as incorrect and the ambition is to replace this incorrect knowledge right from the beginning. From the perspective of the psychology of learning, this is indeed a very large obstacle to be overcome. In addition, this approach is largely mathematics oriented. The objective is to work towards coordinate-free descriptions and in particular to avoid the „coordinate virus” described by Hestenes [Hestenes 1992].

In order not to render mathematical unlearning as the first task when beginning to work with geometric algebra, it does make sense to first focus the attention on the central points of physics, didactically reconstructing geometric algebra. And this central point of physics lies in the interpretation of the Pauli matrices. Therefore it is suggested [Horn 2004], within the context of the description of the classic three-dimensional space, to position the interpretation of Pauli matrices at the beginning of didactically reconstructing geometric algebra.

[5] This word in [Gull et al. 1993] also means „to break a habit“ and thus indicates didactic problems associated with this approach.

ist deshalb nicht die Gleichsetzung mit einer 2x2-Matrizenrepräsentation das Entscheidende, sondern die linksseitig formulierte Beziehung zwischen Quaternionen und Pauli-Matrizen, also einer Beziehung zwischen zwei unterschiedlichen mathematischen Konstrukten.

### *Zur Didaktik der Geometrischen Algebra*

Es bestehen zahlreiche Anknüpfungspunkte, die Geometrische Algebra didaktisch aufzuarbeiten. In den üblichen Darstellungen zur Geometrischen Algebra steht zumeist die Deutung des geometrischen Produkts $\mathbf{ab} = \mathbf{a} \cdot \mathbf{b} + \mathbf{a} \wedge \mathbf{b}$ an Anfang der Behandlung der Geometrischen Algebra.

Begründet wird dieses Vorgehen mit der Tatsache, dass im Rahmen der heute verwendeten Vektoralgebra das Produkt zweier Vektoren inkonsistent und uneindeutig definiert ist. Tatsächlich existieren mit dem Skalarprodukt und dem Kreuzprodukt (Vektorprodukt) zwei unterschiedliche Produkte von Vektoren, die in den Lehrbüchern isoliert voneinander präsentiert werden. Darüber hinaus gestattet das Kreuzprodukt keine eindeutige Verallgemeinerung in höherdimensionalen Räumen.

Dabei wird vom Lernenden zuerst ein „Umlernen“ („a little un-learning“)[5] [Gull et al. 1993] verlangt. Das bisherige Wissen der Studentinnen und Studenten wird als fehlerhaft dargestellt und soll gleich zu Anbeginn ersetzt werden. Dies stellt lernpsychologisch eine sehr hohe Hürde dar. Darüber hinaus ist diese Vorgehen stark mathematikorientiert. Insbesondere soll der von Hestenes beschriebene „Koordinatenvirus“ nicht unterstützt und auf koordinatenfreie Beschreibungen hingearbeitet werden [Hestenes 1992].

Um nicht ein mathematisches Umlernen an den Anfang der Behandlung der Geometrischen Algebra zu stellen, ist es sinnvoll, den physikalischen Kernpunkt ins Zentrum einer didaktischen Rekonstruktion der Geometrischen Algebra zu rücken – und dieser physikalische Kern findet sich in der Interpretation der Pauli-Matrizen. Deshalb wird vorgeschlagen [Horn 2004], die Deutung der Pauli-Matrizen im Kontext der Beschreibung des klassischen dreidimensionalen Raumes als Ausgangspunkt einer didaktischen Rekonstruktion der Geometrischen Algebra zu setzen.

[5] Das im englischen Original in [Gull et al. 1993] verwendete Verb „to unlearn“ bedeutet darüber hinaus „abgewöhnen“ und zeigt die didaktische Problematik dieses Ansatzes.

## 5. Duality

In the following section I will try to interrelate quaternions and geometric algebra in a consistent manner. The interrelationship as such represents nothing new and has already been described in books and review articles of geometric algebra (see for example [Doran et al. 2003, p. 33/34] ). The relationship however will be considered here from another perspective, i.e. from the perspective of duality.

According to Lee Smolin, the principle of duality should receive a similar status as the principle of relativity has in physics, and „it is a important as … the principle of relativity" [Smolin 203, p. 113]. „The principle of duality applies when two descriptions are different ways of looking at the same thing" [Smolin 2003, p. 214].

Moreover „the idea of duality is still a major driving force behind research in elementary particle physics and string theory" [Smolin 2003, p. 117]. Michael Atiyah lists a number of examples that are based on the duality principle, starting with the duality between electricity and magnetism that lead Dirac to postulate magnetic monopoles. Thus, it becomes feasible that in several theoretical descriptions solitons and fields of elementary particles or weak and strong interactions can be perceived as two different perspectives on the same underlying phenomenon [Atiyah 1998, p. 121].

For this reason it is also interesting from a didactic perspective that in geometric algebra a mathematically simple and clear example is available for dual associations. The idea of duality is thus much more accessible to learners without having to overcome larger mathematic obstacles.

### *Duality in Geometric Algebra*

In geometric algebra the eight basic elements can be transformed by pairs into each other with a simple multiplication using the basic trivector $\sigma_x\sigma_y\sigma_z$ (see figure 5). This operation is referred to as a duality transformation[6].

Since it is possible to explicitly associate scalars and trivectors as well as vectors and bivectors, trivectors are also referred to as pseudoscalars and bivectors also as pseudovectors.

[6] Again there are variations in the choice of algebraic signs. The writing style chosen here orients itself at [Doran et al. 2003, p. 35] and is referred to in [Baylis 2002, p. 21] as Clifford dual and conceptually different from Hodge dual (multiplication with the inverse pseudoscalar $\sigma_z\sigma_y\sigma_x$ ).

## 5. Dualität

Im nun folgenden Teil soll versucht werden, Quaternionen und Geometrische Algebra konsistent zu verknüpfen. Diese Verknüpfung ist an sich nichts Neues und wird in Darstellungen der Geometrischen Algebra (siehe beispielsweise [Doran et al. 2003, S. 33/34] ) beschrieben. Der Zusammenhang wird hier jedoch unter einem anderem Blickwinkel – dem Blickwinkel der Dualität – betrachtet.

Lee Smolin zufolge kommt dem Prinzip der Dualität eine ähnliche Bedeutung in der Physik zu wie dem Prinzip der Relativität und „es ist so wichtig ... wie das Relativitätsprinzip" [Smolin 2003, S. 113]. Die Dualität kommt immer dann zur Wirkung, wenn es mehrere verschiedene Möglichkeiten gibt, ein und denselben Sachverhalt zu betrachten [Smolin 2003, S. 214] ).

Im Bereich der Elementarteilchenphysik und der String-Theorie stellt diese Idee „weiterhin eine Hauptantriebskraft" [Smolin 2003, S. 117] dar. Angefangen mit der Dualität zwischen Elektrizität und Magnetismus, die Dirac zur Postulierung von magnetischen Monopolen führte, zählt Michael Atiyah eine ganze Reihe von Beispielen auf, die auf das Dualitätsprinzip zurück gehen. So können in einigen theoretischen Beschreibung Solitonen und Elementarteilchenfelder oder die schwache und die starke Wechselwirkung als zwei unterschiedliche Blickweisen auf das gleiche zugrunde liegende Phänomen betrachtet werden [Atiyah 1998, S. 121].

Es ist deshalb auch didaktisch interessant, dass in der Geometrischen Algebra ein mathematisch sehr einfaches Beispiel für einen dualen Zusammenhang zur Verfügung steht. Lernende können dadurch an die Idee der Dualität herangeführt werden, ohne größere mathematische Hürden überwinden zu müssen.

### *Dualität in der Geometrischen Algebra*

In der Geometrischen Algebra lassen sich die acht Basisgrößen durch eine einfache Multiplikation mit dem Basis-Trivektor $\sigma_x\sigma_y\sigma_z$ paarweise ineinander überführen (siehe Abbildung 5). Diese Transformation wird als Dualbildung[6] bezeichnet.

Da dadurch Skalare und Trivektoren, sowie Vektoren und Bivektoren eindeutig zugeordnet werden können, werden Trivektoren auch als Pseudoskalare und Bivektoren auch als Pseudovektoren bezeichnet.

[6] Auch dabei gibt es wieder Variationen bei der Vorzeichenwahl. Die hier gewählte Schreibweise orientiert sich an [Doran et al. 2003, S. 35] und wird in [Baylis 2002, S. 21] als Clifford-Dual bezeichnet und vom Hodge-Dual (Multiplikation mit dem inversen Pseudoskalar $\sigma_z\sigma_y\sigma_x$) begrifflich unterschieden.

The duality factor $\sigma_x\sigma_y\sigma_z$ plays a significant role in geometric algebra as the unit volume element, since it commutes with all other possible multivectors in geometric algebra.

The usual abbreviation $\mathbf{I} = \sigma_x\sigma_y\sigma_z$ also reveals with its writing style the algebraic fact that we are dealing with an imaginary parameter, because $\mathbf{I}^2 = -1$. The duality transformation can also be interpreted as a „complexification“[7] with fascinating conceptual consequences: the three-dimensional Euclidean space (within which we lived before Einsteins discovery of the theory of relativity) has a naturally complex structure!

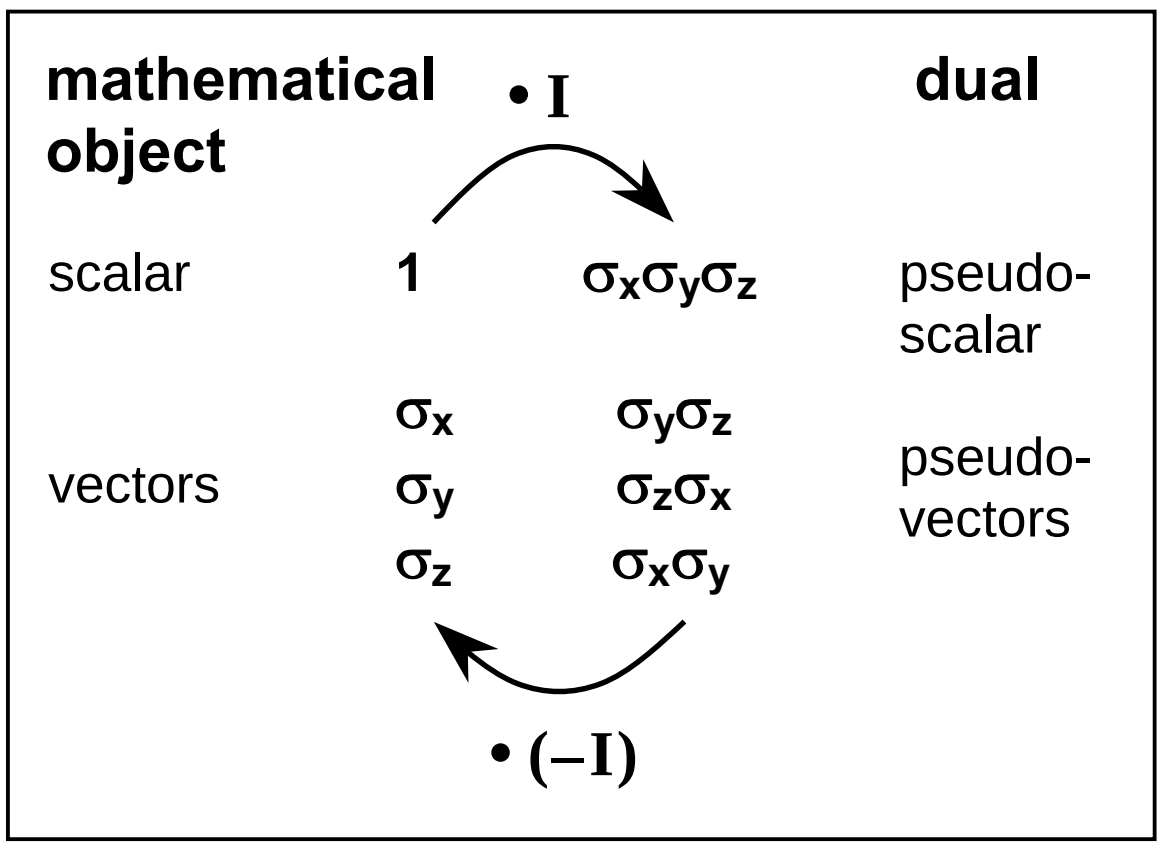

Fig. 5: Dual of basic elements of geometric algebra.

***The Interpretation of Quaternions as Duals***

The algebraic Jordan-Cartan formulae

$$\boldsymbol{i} = -\,\mathbf{i}\,\sigma_x$$
$$\boldsymbol{j} = -\,\mathbf{i}\,\sigma_y$$
$$\boldsymbol{k} = -\,\mathbf{i}\,\sigma_z$$

now allow a second, a dual and especially a geometric view of the basic parameters of quaternions. If one identifies the complex unity **i** with the pseudoscalar **I**, then they can be interpreted as a result of a duality transformation within the context of geometric algebra:

$$\boldsymbol{i} = -\,\mathbf{I}\,\sigma_x = -\,\sigma_x\sigma_y\sigma_z\,\sigma_x = -\,\sigma_y\sigma_z$$
$$\boldsymbol{j} = -\,\mathbf{I}\,\sigma_y = -\,\sigma_x\sigma_y\sigma_z\,\sigma_y = -\,\sigma_z\sigma_x$$
$$\boldsymbol{k} = -\,\mathbf{I}\,\sigma_z = -\,\sigma_x\sigma_y\sigma_z\,\sigma_z = -\,\sigma_x\sigma_y$$

The purely complex part of a quaternion can thus be interpreted as a bivector, i.e. as a directed plane. Or formulated in other words: if Pauli matrices are vectors then pure quaternions are planes.

[7] This approach is implemented by Baylis for constructing paravectors [Baylis 2002].

Der vermittelnde Faktor $\sigma_x\sigma_y\sigma_z$ spielt in der Geometrischen Algebra als Einheits-Volumenelement eine wesentliche Rolle, da er mit allen anderen Größen in der Geometrischen Algebra kommutiert.

Die übliche Abkürzung $\mathbf{I} = \sigma_x\sigma_y\sigma_z$ macht durch ihre Schreibweise auch die algebraische Tatsache deutlich, dass es sich dabei um eine imaginäre Größe handelt, denn $\mathbf{I}^2 = -1$. Die Dualbildung lässt sich also auch als „Komplexifizierung“ deuten[7] – mit faszinierenden konzeptuellen Konsequenzen: Der dreidimensionale euklidische Raum (in dem wir vor Einsteins Entdeckung der Relativitätstheorie lebten) hat eine natürliche komplexe Struktur!

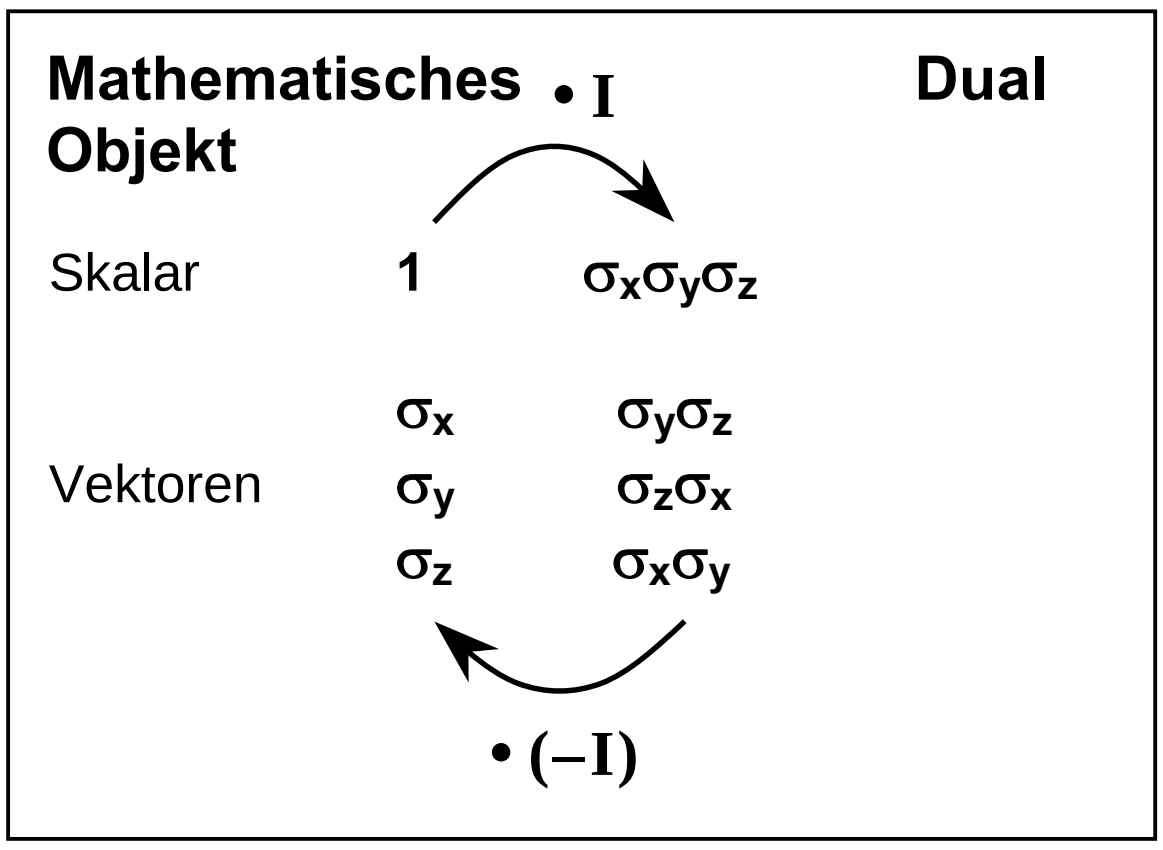

Abb. 5: Dual der Basisgrößen der Geometrischen Algebra.

***Interpretation der Quaternionen als Duale***

Die algebraischen Jordan-Cartanschen Formeln

$$\boldsymbol{i} = -\,\mathbf{i}\,\sigma_x$$
$$\boldsymbol{j} = -\,\mathbf{i}\,\sigma_y$$
$$\boldsymbol{k} = -\,\mathbf{i}\,\sigma_z$$

gestatten nun einen zweiten, dualen und vor allem geometrischen Blick auf die Basisgrößen der Quaternionen. Identifiziert man die komplexe Einheit **i** mit dem Pseudoskalar **I**, lassen sie sich im Kontext der Geometrischen Algebra als Dualbildung interpretieren:

$$\boldsymbol{i} = -\,\mathbf{I}\,\sigma_x = -\,\sigma_x\sigma_y\sigma_z\,\sigma_x = -\,\sigma_y\sigma_z$$
$$\boldsymbol{j} = -\,\mathbf{I}\,\sigma_y = -\,\sigma_x\sigma_y\sigma_z\,\sigma_y = -\,\sigma_z\sigma_x$$
$$\boldsymbol{k} = -\,\mathbf{I}\,\sigma_z = -\,\sigma_x\sigma_y\sigma_z\,\sigma_z = -\,\sigma_x\sigma_y$$

Der rein komplexe Anteil eines Quaternions kann somit als Bivektor, also als orientierte Fläche, interpretiert werden. Oder anders formuliert: Wenn Pauli-Matrizen Vektoren sind, dann sind reine Quaternionen Flächen.

[7] Dieser Ansatz wird von Baylis zur Konstruktion von Paravektoren [Baylis 2002] genutzt.

Real quaternions

$$x = x_0 + x_1 \boldsymbol{i} + x_2 \boldsymbol{j} + x_3 \boldsymbol{k} \qquad (x_i \in \boldsymbol{R})$$

can also feature a scalar proportion $x_0$, so that we come up with the following sum:

$$x = x_0 - x_1 \boldsymbol{\sigma_y \sigma_z} - x_2 \boldsymbol{\sigma_z \sigma_x} - x_3 \boldsymbol{\sigma_x \sigma_y}$$

From a geometric view this is the sum of a dimensionless quantity and an oriented plane.

***Back to Complex Numbers***

The three-dimensional Euclidean space with $(\boldsymbol{\sigma_x\sigma_y\sigma_z})$, $(\boldsymbol{\sigma_x\sigma_y})$, $(\boldsymbol{\sigma_y\sigma_z})$ und $(\boldsymbol{\sigma_z\sigma_x})$ possesses four imaginary basic entities that can be squared to **–1**. It behaves differently in the case of a space with only two dimensions. Here only one imaginary basic entity exists: $\boldsymbol{\sigma_x\sigma_y}$. This bivector can be clearly identified therefore (except the algebraic sign) with the complex basic unit **i**.

However, within the context of geometric algebra the imaginary basic element $\boldsymbol{\sigma_x\sigma_y}$ is not a longitudinal line, but rather a directed plane. In order to represent the real and imaginary coordinate axis in figures 1 and 2 physically in a meaningful manner as directed lines and not without dimensionless (real axis) or as a plane (imaginary axis), it is necessary to modify the representation of the axes.

A consistent interpretation of the axes takes place when for example the axes in figure 1 and 2 are multiplied with $\boldsymbol{\sigma_x}$ from the left (see figure 6).

Ein beliebiges reelles Quaternion

$$x = x_0 + x_1 \boldsymbol{i} + x_2 \boldsymbol{j} + x_3 \boldsymbol{k} \qquad (x_i \in \boldsymbol{R})$$

kann auch einen skalaren Anteil $x_0$ aufweisen, so dass sich insgesamt ergibt:

$$x = x_0 - x_1 \boldsymbol{\sigma_y \sigma_z} - x_2 \boldsymbol{\sigma_z \sigma_x} - x_3 \boldsymbol{\sigma_x \sigma_y}$$

Geometrisch stellt dies die Summe aus einer dimensionslosen Größe und einer orientierten Fläche dar.

***Zurück zu den komplexen Zahlen***

Der dreidimensionale euklidische Raum besitzt mit $(\boldsymbol{\sigma_x\sigma_y\sigma_z})$, $(\boldsymbol{\sigma_x\sigma_y})$, $(\boldsymbol{\sigma_y\sigma_z})$ und $(\boldsymbol{\sigma_z\sigma_x})$ vier imaginäre Basisgrößen, die sich zu **–1** quadrieren. Anders verhält es sich im zweidimensionalen Fall. Hier gibt es nur die eine imaginäre Basisgröße $\boldsymbol{\sigma_x\sigma_y}$. Diese lässt sich deshalb (bis auf das Vorzeichen) eindeutig mit der komplexen Basiseinheit **i** identifizieren.

Allerdings stellt die imaginäre Basisgröße $\boldsymbol{\sigma_x\sigma_y}$ im Kontext der Geometrischen Algebra keine längenartige Größe, sondern eine orientierte Fläche dar. Um die reelle und imaginäre Koordinatenachse in den Abbildungen 1 und 2 physikalisch sinnvoll längenartig und nicht als dimensionslos (reelle Achse) bzw. flächenartig (imaginäre Achse) darzustellen, ist eine Modifikation der Achsendarstellung notwendig.

Eine konsistente Interpretation der Achsen ergibt sich, indem beispielsweise die Achsen in Abbildung 1 und 2 *linksseitig* mit $\boldsymbol{\sigma_x}$ multipliziert werden (siehe Abbildung 6).

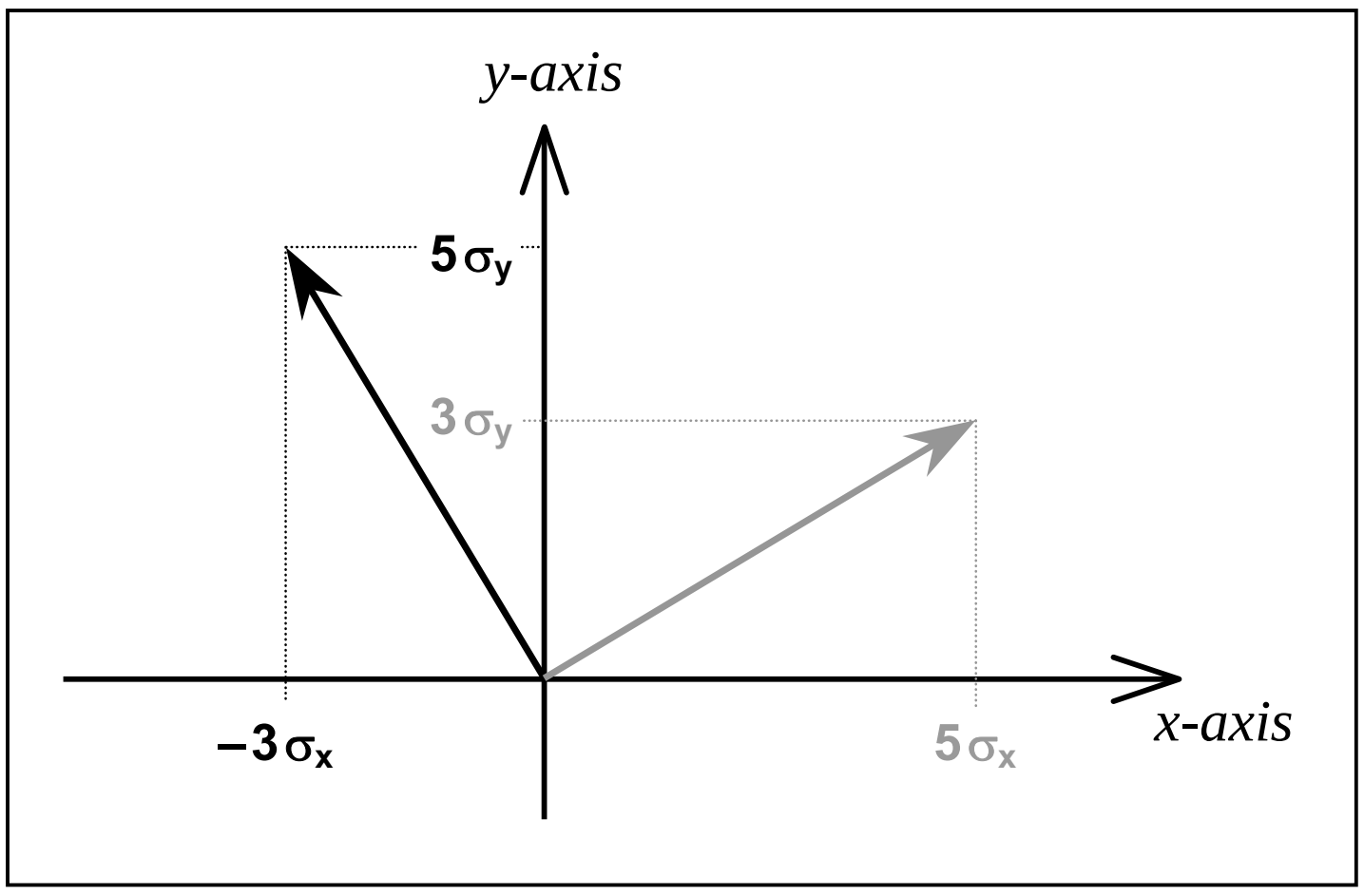

Fig. 6: Modification of the complex plane in geometric algebra.

Abb. 6: Modifikation der Gauß'schen Zahlenebene in der Geometrischen Algebra.

A generalisation of the interpretation of $\boldsymbol{\sigma_x\sigma_y}$ as an operator that induces a rotation of 90° (which is the angle between the vectors $\boldsymbol{\sigma_y}$ and $\boldsymbol{\sigma_x}$) can be constructed for higher-dimensional spaces (see [Hestenes 1990], [Hestenes 2002], [Doran et al. 2003] and other descriptions of geometric algebra). Due to the non-commutativity, it is necessary to maintain consistency when multiplying from the left and when multiplying from the right.

Die Interpretation von $\boldsymbol{\sigma_x\sigma_y}$ als ein Operator, der eine Rotation um den zwischen den Vektoren $\boldsymbol{\sigma_y}$ und $\boldsymbol{\sigma_x}$ liegenden Winkel von 90° vermittelt, lässt sich auf höherdimensionale Räume verallgemeinern (siehe [Hestenes 1990], [Hestenes 2002], [Doran et al. 2003] und andere Darstellungen der Geometrischen Algebra). Zu berücksichtigen ist aufgrund der Nichtkommutativität dabei jedoch, dass konsistent mit rechts- und linksseitiger Multiplikation umgegangen wird.

## 6. Complex Quaternions

Quaternions with complex coefficients

$$z = z_0 + z_1\boldsymbol{i} + z_2\boldsymbol{j} + z_3\boldsymbol{k} \qquad (x_i \in \boldsymbol{C})$$

are referred to as biquaternions. A diversity of controversial views exists with regard to whether these expanded quaternions are allegedly useful or useless. While Anderson and Joshi, as quoted above, attribute an important heuristic function especially to complex quaternions and numerous attempts are made in reformulating physical relationships with the help of complex quaternions still today [Hans 2006], other authors exhibit partly ambivalent and conflicting opinions in this regard.

Paul A. M. Dirac for instance recognises that it might be possible to relate the physical world to biquaternions: „In this way the physical world is put into correspondance with the scheme of bi-quarternions, instead of with the scheme of quaternions." But as he continues: „Now the scheme of bi-quarternions is not of any special interest in mathematical theory…", he concludes that this scheme is no adequate tool for the development of physical theories [Dirac 1945, p. 261].

Diracs view is not only based on his conviction of the „relevance of beauty" in mathematical physics [Fischer 1997, p. 44], but his view is also based on his concept concerning the mathematical modelling of our world. „Also, they have eight components, which is rather too many for a simple scheme for describing quantities in space-time" [Dirac 1945, p. 261]. In this regard he is not entirely right – there are $2^3$ components and thus not enough for a complete treatment of space-time in four dimensions. This however is the perfect number of components for describing our three-dimensional, pre-relativistic world.

### *Complex Quaternions and Geometric Algebra*

With the concept of a duality transformation the eight basic elements of geometric algebra can be linked with the eight components of complex quaternions. Based on the interrelationships illustrated in figure 5, every real quaternion

$$y = y_0 - y_1\boldsymbol{\sigma_y\sigma_z} - y_2\boldsymbol{\sigma_z\sigma_x} - y_3\boldsymbol{\sigma_x\sigma_y}$$

can be transformed into the affiliated dual which then is

$$\mathbf{I}\,y = y_1\boldsymbol{\sigma_x} + y_2\boldsymbol{\sigma_y} + y_3\boldsymbol{\sigma_z} + y_0\boldsymbol{\sigma_x\sigma_y\sigma_z}$$

This expression can be identified with the complex part of a biquaternion so that a complex quaternion $z = x + \mathbf{i}\,y$ can be clearly designated with the following multivector in geometric algebra:

## 6. Komplexe Quaternionen

Quaternionen mit komplexen Koeffizienten

$$z = z_0 + z_1\boldsymbol{i} + z_2\boldsymbol{j} + z_3\boldsymbol{k} \qquad (x_i \in \boldsymbol{C})$$

werden als Biquaternionen bezeichnet. Über den vermeintlichen Nutzen oder die vermeintliche Nutzlosigkeit dieser erweiterten Quaternionen gibt es kontroverse Ansichten. Während Anderson und Joshi wie oben zitiert gerade den komplexen Quaternionen eine wichtige heuristische Funktion zuordnen und auch heute noch zahlreiche Versuche einer Umformulierung physikalischer Beziehungen mit Hilfe komplexer Quaternionen unternommen werden [Hans 2006], zeigt sich bei anderen Autoren auch hier wieder ein ambivalenter und zwiespältiger Umgang.

So erkennt Paul A. M. Dirac zwar an, dass „die physikalische Welt mit Biquaternionen in Beziehung gesetzt werden kann". Da das Schema der Biquaternionen jedoch in der mathematischen Theoriebildung „überhaupt nicht von besonderem Interesse sei", seien sie ebenfalls nicht zur physikalischen Theoriebildung geeignet [Dirac 1945, S. 261].

Dieser Einschätzung Diracs liegt nicht nur seine Überzeugung von der „Relevanz des Schönen" in der mathematischen Physik zugrunde [Fischer 1997, S. 44], sondern auch sein Konzept bezüglich der mathematischen Modellierung unserer Welt. „Komplexe Quaternionen haben", so schreibt er weiter, „acht Komponenten, was etwas zu viel für ein einfaches Schema sei, um Größen in der Raumzeit zu beschreiben." [Dirac 1945, S. 261]. Da hat er nicht ganz recht – es sind $2^3$ Komponenten und damit etwas zu wenig für eine vollständige Behandlung der Raumzeit mit vier Dimensionen. Aber es ist genau die richtige Anzahl an Komponenten, um unsere dreidimensionale, vorrelativistische Welt zu beschreiben.

### *Komplexe Quaternionen und Geometrische Algebra*

Mit dem Konzept der Dualbildung können die acht Basisgrößen der Geometrischen Algebra mit den acht Komponenten komplexer Quaternionen verknüpft werden. Ausgehend von den in Abbildung 5 gezeigten Zusammenhängen lautet der von einem reellen Quaternion

$$y = y_0 - y_1\boldsymbol{\sigma_y\sigma_z} - y_2\boldsymbol{\sigma_z\sigma_x} - y_3\boldsymbol{\sigma_x\sigma_y}$$

abgeleitete Dual

$$\mathbf{I}\,y = y_1\boldsymbol{\sigma_x} + y_2\boldsymbol{\sigma_y} + y_3\boldsymbol{\sigma_z} + y_0\boldsymbol{\sigma_x\sigma_y\sigma_z}$$

Dies lässt sich mit dem komplexen Anteil eines Biquaternions identifizieren, so dass ein komplexes Quaternion $z = x + \mathbf{i}\,y$ eineindeutig folgendem Multivektor der Geometrischen Algebra zugeordnet werden kann:

$$z = x_0 + y_1\sigma_x + y_2\sigma_y + y_3\sigma_z - x_1\sigma_y\sigma_z - x_2\sigma_z\sigma_x - x_3\sigma_x\sigma_y + y_0\sigma_x\sigma_y\sigma_z \quad (x_i, y_i \in \mathbb{C})$$

Already Élie Cartan pointed at the algebra of complex quaternions: „Any element of the algebra is uniquely the sum of a real scalar, a real vector, a real bivector and a real trivector“ [Cartan 1981, p. 45]. But while Cartan however focused primarily on the algebraic connections, geometric algebra discloses geometric structures (see figure 7).

| geometric algebra | complex quaternions | interpretation |
|---|---|---|
| 1 | 1 | dimensionless number (scalar) |
| $\sigma_x$ | i $\boldsymbol{i}$ | oriented lines (vectors) |
| $\sigma_y$ | i $\boldsymbol{j}$ | |
| $\sigma_z$ | i $\boldsymbol{k}$ | |
| $\sigma_x\sigma_y$ | $-\boldsymbol{k}$ | oriented planes (bivectors) |
| $\sigma_y\sigma_z$ | $-\boldsymbol{i}$ | |
| $\sigma_z\sigma_x$ | $-\boldsymbol{j}$ | |
| $\sigma_x\sigma_y\sigma_z$ | i | oriented volume (trivector) |

Fig. 7: Isomorphism of geometric algebra and algebra of complex quaternions.

**7. Didactical possibilities**

After the publication of his Ausdehnungslehre (Calculus of Extension) Grassmann was confronted with the accusation of having no ability to visualize his theory. „Such an *abstract* calculus of extension, as he is searching for, can only be developed from concepts. However, the source of mathematical cognition does not lie in concepts, but in the vision and visualization,“ Apelt remarked in a letter to Möbius in 1845 (quoted according to [Petsche 2006, p. XII). The underlying direction of action at that time is clear: from vision to abstraction.

The current notion of mathematics seems to be diametrically opposed to the conceptions that were predominant at that time. Roger Penrose calls his book „The Road to Reality” which deals with the mathematical principles of physics, and the message is clear: without establishing abstract principles one cannot obtain a sensible view of reality [Penrose 2005].

Our description of the world, of which mathematics is an integral part, has come a long way since

$$z = x_0 + y_1\sigma_x + y_2\sigma_y + y_3\sigma_z - x_1\sigma_y\sigma_z - x_2\sigma_z\sigma_x - x_3\sigma_x\sigma_y + y_0\sigma_x\sigma_y\sigma_z \quad (x_i, y_i \in \mathbb{C})$$

Auf die Algebra komplexer Quaternionen hat schon Élie Cartan hingewiesen: „Jedes Element dieser Algebra ist die eineindeutige Summe eines reellen Skalars, eines reellen Vektors, eines reellen Bivektors und eines reellen Trivektors.“ [Cartan 1981, S. 45]. Doch während Cartan hauptsächlich die algebraischen Zusammenhänge im Blick hatte, legt die Geometrische Algebra die geometrischen Strukturen offen (siehe Abbildung 7).

| Geometr. Algebra | Komplexe Quaternionen | Interpretation |
|---|---|---|
| 1 | 1 | dimensionslose Größe (Skalar) |
| $\sigma_x$ | i $\boldsymbol{i}$ | gerichtete Strecken (Vektoren) |
| $\sigma_y$ | i $\boldsymbol{j}$ | |
| $\sigma_z$ | i $\boldsymbol{k}$ | |
| $\sigma_x\sigma_y$ | $-\boldsymbol{k}$ | orientierte Flächen (Bivektoren) |
| $\sigma_y\sigma_z$ | $-\boldsymbol{i}$ | |
| $\sigma_z\sigma_x$ | $-\boldsymbol{j}$ | |
| $\sigma_x\sigma_y\sigma_z$ | i | orientiertes Volumen (Trivektor) |

Abb. 7: Isomorphie von Geometrischer Algebra und Algebra der komplexen Quaternionen.

**7. Didaktische Möglichkeiten**

Nach Publikation seiner Ausdehnungslehre musste sich Graßmann noch mit dem Vorwurf mangelnder Anschauung auseinander setzen. „So eine *abstrakte* Ausdehnungslehre, wie er sucht, könne sich nur aus Begriffen entwickeln lassen. Aber die Quelle der mathematischen Erkenntnis liegt nicht in Begriffen, sondern in der Anschauung.“ bemerkte Apelt in einem Brief an Möbius 1845 (zitiert nach [Petsche 2006, S. XII] ). Die damals zugrunde gelegte Wirkungsrichtung ist klar: Von der Anschauung zur Abstraktion.

Die heutige Mathematikauffassung scheint den damals vorherrschenden Vorstellungen diametral entgegen gesetzt zu sein. „The Road to Reality“ – Den Weg zur Wirklichkeit nennt Roger Penrose sein Buch über die mathematischen Grundlagen der Physik, und die Botschaft ist klar: Ohne abstrakte Grundlagensetzung ist keine vernünftige Sicht auf die Realität möglich [Penrose 2005].

Unsere Weltbeschreibung, die die Mathematik als einen wesentlichen Bestandteil einschließt, hat seit

Grassmann. It does not consider vision or visualization, but the abstract-axiomatic basis as the true foundation of science.

Between these two – historic – poles it seems as though the framework of modern didactic approaches in physics acts in a mediating and integral manner. On the one hand one can describe „the becoming of physical concepts as being based on phenomenon“ [Mikelskis 2006, p. 98], where the vision is located at the beginning of the learning process. On the other hand, observations in the natural sciences are based on theories [Mikelskis 2006, p. 124], and this dependence on theory is considered to be an important aspect of physics education.

The debate on theory versus vision is being mitigated didactically within the framework of a geometric-dual interpretation of quaternions. Quaternions can on the one hand become more accessible by their clear interpretation. In this case at the starting point of such a didactic approach are not formal and abstract relationships, but rather lines, planes and volumes.

On the other hand, learners gain access to the fact that physical and mathematical interpretations are always based on theories. Depending on the subjective standpoint taken by learners or teachers, either the structural-geometric, the dynamic-operational or the formal-abstract aspects of quaternions can emerge. By means of this interpretative power of quaternionic entities, students can discover that theory and reality do not built on each other in a linear manner, but are inherently entangled and interwoven with each other.

***Metaconceptual Competence***

Not only the development of modelling skills is considered to be a significant competence, but also the comparison of different models. This comparing and categorizing of different models is a skill that should be supported when learning physics [Mikelskis 2006, p. 120 ff]. As already described above, the classification and model-oriented analysis of quaternions can lead to expanding metaconceptual competences.

Such a metaconceptual competence has definite advantages as concerns learning. Learners with a metaconceptual understanding are thus not only able to make a deliberate and problem-oriented change from patterns of reasoning of one model into the level of another model, but are furthermore capable to name as well as explain these changes and to describe the borders of each particular model.

On the other hand, metaconceptual competence enables learners to adequately and competently deal

Graßmann einen langen Weg zurückgelegt. Sie sieht nicht mehr die Anschauung, sondern die abstrakt-axiomatische Basis als wahres Fundament der Wissenschaft an.

Zwischen diesen beiden – historischen – Polen scheint der Handlungsrahmen moderner physikdidaktischer Ansätze vermittelnd und integrierend zu wirken. Einerseits wird „das Werden physikalischer Begriffe ausgehend von den Phänomenen“ [Mikelskis 2006, S. 98] beschrieben, in der die Anschauung am Anfangspunkt des Lernprozesses steht. Andererseits wird die Theoriegeladenheit naturwissenschaftlicher Beobachtungen [Mikelskis 2006, S. 124] als wesentlicher Teil physikdidaktischer Überlegungen anerkannt.

Das Spannungsfeld zwischen Theorie und Anschauung wird im Rahmen einer geometrisch-dualen Interpretation der Quaternionen didaktische entschärft. Einerseits können die Quaternionen einer anschaulichen Interpretation zugänglich gemacht werden. Am Ausgangspunkt eines solchen didaktischen Ansatzes stehen dann nicht formale, abstrakte Beziehungen, sondern Strecken, Flächen und Volumina.

Andererseits kann mit den Lernenden die Theoriegeladenheit physikalischer und mathematischer Interpretationen erarbeitet werden. Je nach subjektiv eingenommenem Standpunkt kann der flächenhaft-geometrische, der dynamisch-operationelle oder der formal-abstrakte Aspekt der Quaternionen hervor treten. Anhand dieser Interpretationsmächtigkeit quaternionischer Größen können Studierende feststellen, dass Theorie und Wirklichkeit nicht linear aufeinander aufbauen, sondern inhärent miteinander verschränkt und verwoben sind.

***Metakonzeptuelle Kompetenz***

Nicht nur das Erarbeiten von Modellierungen, sondern das Nebeneinanderstellen und Vergleichen unterschiedlicher Modelle stellt eine wesentliche Kompetenz dar, die Lernenden vermittelt werden sollte [Mikelskis 2006, S. 120 ff]. Wie oben bereits beschrieben, kann die Einordnung und modellorientierte Analyse der Quaternionen zur Ausbildung einer solchen metakonzeptuellen Kompetenz führen.

Diese metakonzeptuelle Kompetenz hat einerseits deutlich lernbezogene Vorteile. So sind Lernende mit einem metakonzeptuellen Verständnis nicht nur in der Lage, bewusst und problemorientiert von Argumentationsmustern eines Modells in die Ebene eines anderen Modells zu wechseln, sondern darüber hinaus fähig, diese Wechsel zu benennen, zu begründen und die Grenzen der jeweiligen Modelle zu beschreiben.

Andererseits befähigt die metakonzeptuelle Kompetenz die Lernenden, sachangemessen mit erkenntnis-with epistemological questions [Horn 2006. Where is the line drawn at the border between the real world and the model world, between reality and actuality in our minds? Is for example the 4π-symmetry in modern physics an artificial artefact that is produced as a result of the selection of the model or does it express a fundamental physical property?

***Education and Duality***

Properties and the interpretation of quaternions can be understood using the concept of duality. This however is no one-sided process in only one direction. The didactical effect can also be oriented into the opposite direction: quaternions help illustrate and exemplify the concept of duality.

Since duality takes on a central role in the history of ideas in modern physics, the teaching of the concept of duality is of extreme importance at the university level and in instruction in general. The manner in which the concept is handled and presented didactically in physics however is still in its beginning stages.

***The Nature of Mathematics***

Mathematics is, as a German dictionary [Drosdowski et al. 1996] succinctly notes, the science of space and numbers. The two corresponding fields of algebra and geometry, however, are mostly taught independently and remain unconnected in seminars and lectures at the university and in schools. This is also unfortunate in terms of physics education and teaching.

The following is meanwhile true what David Hestenes notes in an often quoted statement: „Geometry without algebra is dumb! – Algebra without geometry is blind!“ (quoted according to [Gull et al. 1993] ) The at times simultaneous dumbness and blindness in the field of mathematics can be overcome by means of geometric algebra. Hestenes thus gives geometry a language and algebra a face.

## 8. Outlook

What this language and face may look like can for instance be demonstrated by determining the square roots within the context of geometric algebra. The only thing that then needs to be done is to translate the particular results from complex quaternions [Hans 2005], [Sangwine 2005] to geometric algebra.

As a result of such a procedural method, the construct of the square root receives a supplementary, geometric-spatial meaning.

Spaces of higher dimensions can be described using spacetime algebra, an expanded version of geometric algebra [Hestenes 1986]. Taking the different signatures of spatial and temporal coordinates into consideration, then Dirac matrices constitute the theoretischen Fragestellungen [Horn 2006] umzugehen. Wo wird die Grenzlinie zwischen Realwelt und Modellwelt, zwischen Realität und Wirklichkeit gezogen? Ist beispielsweise die 4 π-Periodizität in der modernen Physik ein künstliches, durch die Modellwahl erzeugtes Artefakt oder drückt sich damit eine grundlegende physikalische Eigenschaft aus?

***Didaktik und Dualität***

Mit Hilfe des Konstrukts der Dualität können Eigenschaften und Interpretation von Quaternionen behandelt werden. Dies ist jedoch in der Wirkungsrichtung kein einseitiger Vorgang, denn es gilt auch umgekehrt: Mit Hilfe der Quaternionen kann der Begriff der Dualität veranschaulicht und exemplarisch dargestellt werden.

Da die Dualität in der Ideengeschichte der modernen Physik einen zentralen Platz einnimmt, ist die Vermittlung der Dualität in Studium und Unterricht entsprechend wichtig. Die physikdidaktische Aufarbeitung des Begriffs der Dualität steht jedoch noch am Anfang.

***Die Natur der Mathematik***

Die Mathematik ist, wie der Duden [Drosdowski et al. 1996] knapp vermerkt, die Wissenschaft von den Raum- und Zahlengrößen. Die Themenfelder von Algebra und Geometrie stehen jedoch im universitären und schulischen Vermittlungsprozess meist unverknüpft nebeneinander. Dies ist auch aus didaktischen Gründen bedauerlich.

Denn es gilt, was David Hestenes in einem mittlerweile oft zitierten Ausdruck feststellt: „Geometrie ohne Algebra ist stumm! – Algebra ohne Geometrie ist blind!“ (zitiert nach [Gull et al. 1993] ) Mit Hilfe der Geometrischen Algebra kann die heute zuweilen herrschende gleichzeitige Stummheit und Blindheit im Bereich der Mathematik überwunden werden. Hestenes gibt damit der Geometrie eine Sprache und der Algebra ein Gesicht.

## 8. Ausblick

Wie diese Sprache und dieses Gesicht aussieht, kann beispielsweise bei der Bestimmung der Quadratwurzeln im Kontext der Geometrischen Algebra gezeigt werden. Es sind dabei lediglich die entsprechenden Ergebnisse aus dem Bereich der komplexen Quaternionen [Hans 2005], [Sangwine 2005] in die Geometrische Algebra zu übersetzen.

Durch eine solche Vorgehensweise erhält das Konstrukt der Quadratwurzel eine ergänzende, geometrisch-räumliche Interpretationsrichtung.

Auch Räume höherer Dimension lassen sich mit Hilfe der Raumzeit-Algebra, einer erweiterten Geometrischen Algebra, beschreiben [Hestenes 1986]. Berücksichtigt man die unterschiedliche Signatur räumlicher und zeitlicher Größen, so bilden die vectorial base of the four-dimensional pseudo-Euclidean space-time.

What is even more important for teaching modern physics is the possibility to clearly illustrate simple conformal structures. The five-dimensional conformal geometric algebra, as it is already used within the framework of computer graphics [Hildenbrand 2005], provides an easily understandable connecting point. It is here where it becomes necessary to process and utilise this point of contact in physics education.

Dirac-Matrizen die vektorielle Basis der vierdimensionalen pseudoeuklidischen Raumzeit.

Für die Vermittlung der modernen Physik noch bedeutsamer ist die Möglichkeit, einfache konformale Strukturen anschaulich darzustellen. Die fünfdimensionale konformale Geometrische Algebra, wie sie heute schon im Rahmen der Computergraphik verwendet wird [Hildenbrand 2005], bietet hier einen leicht verständlichen Anknüpfungspunkt. Den gilt es, auch physikdidaktisch aufzuarbeiten und physikdidaktisch zu nutzen.

## 9. Literatur